\documentclass[aps,prd,nofootinbib,preprintnumbers,12pt,
tightenlines,superscriptaddress]{revtex4}

\usepackage{graphicx}
\usepackage{rotating}
\usepackage{latexsym}

\newcommand{\beq}{\begin{equation}}
\newcommand{\eeq}{\end{equation}}
\newcommand{\bqa}{\begin{eqnarray}}
\newcommand{\eqa}{\end{eqnarray}}

\begin{document}

\preprint{BI-TP 2006/27}

\title{Momentum Broadening in an Anisotropic Plasma}

\author{Paul Romatschke}
\affiliation{Fakult\"at f\"ur Physik, Universit\"at Bielefeld,
D-33501 Bielefeld, Germany}
\date{\today}

\begin{abstract}
The rates governing momentum broadening in a quark-gluon plasma with a 
momentum anisotropy are calculated to leading-log order 
for a heavy quark using kinetic
theory. It is shown how the problematic singularity for these rates at
leading-oder is lifted by next-to-leading order gluon self-energy corrections to give a finite
contribution to the leading-log result.
The resulting rates are shown to lead to larger momentum broadening
along the beam axis than in the transverse plane, which is consistent
with recent STAR results.
This might indicate that the quark-gluon-plasma at RHIC is not in
equilibrium. 
\end{abstract}

\maketitle

\section{Introduction}

One of the hottest debated questions in the context of
ultrarelativistic heavy ion collisions is whether a 
``thermal'' quark-gluon plasma has been created at the highest energy runs of the Relativistic
Heavy-Ion Collider (RHIC).
On the one hand, 
the success of ideal hydrodynamic fits to experimental data have been
interpreted \cite{Heinz:2005zg} to imply rapid thermalization of the
bulk matter at RHIC. 
On the other hand, perturbative estimates of the thermalization time
\cite{Baier:2000sb} result in much larger values, seemingly excluding the
possibility of rapid thermalization (see however \cite{Xu:2004mz}) .
Plasma instabilities 
may help speeding up the equilibration process
by rapidly isotropizing the system
\cite{Mrowczynski:1993qm,Mrowczynski:2005ki,Arnold:2004ti,Rebhan:2004ur,Romatschke:2005pm},
probably by some cascade to the UV
\cite{Arnold:2005ef,Arnold:2005qs,Dumitru:2006pz,Romatschke:2006nk,
Mueller:2006up}, 
although they might still be too slow to overcome the initially strong
longitudinal expansion rate \cite{Romatschke:2006wg}.
Thus, it is currently unclear how (and if) the quark gluon plasma 
created at RHIC stays locally isotropic (let alone thermal) during most
of its time evolution.

Nevertheless, many theoretical calculations in the context of
heavy-ion collisions assume a homogeneous and locally isotropic
system. However, the physics of anisotropic plasmas (even if they are
close to isotropy) can be quite different from that of isotropic
plasmas, c.f. the presence of plasma instabilities in the former.
Due to these potentially large differences, 
it might be interesting to reanalyze calculations of
experimentally accessible observables in the context of anisotropic
plasmas, wherever possible.

In what follows, I will thus study momentum broadening in a homogeneous
but locally anisotropic system. Since I will be interested in
qualitative effects only, I limit myself to considering momentum
broadening for a heavy quark induced by collisions to 
leading logarithm (LL) accuracy. 
In section \ref{sec:setup}, I will discuss the calculational setup for anisotropic plasmas.
Section \ref{sec:mombroad} will contain the calculation of the rates governing momentum 
broadening, which can be done analytically in the small anisotropy limit. 
In section 
\ref{sec:discussion} finally I will try to apply these results to preliminary STAR data 
on jet shapes and discuss possible consequences.

\section{Setup}
\label{sec:setup}

The setup of my calculation will be based on the work by Moore and
Teaney \cite{Moore:2004tg}. They considered a heavy quark 
with momentum $p$ moving in a static (thermal) medium with a
temperature $T$ and I will quickly repeat some equations from their
work which I want to use in the following.

\begin{figure}
\begin{center}
\includegraphics[width=0.5\linewidth,viewport=0 400 575 700]{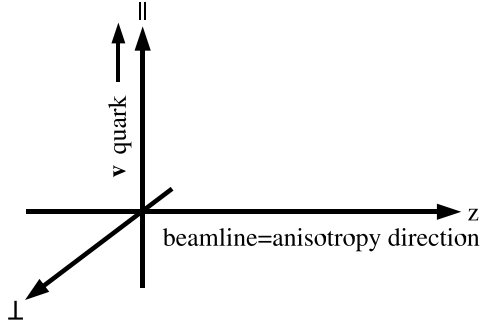}
\end{center}
\caption{
Sketch of the coordinate system: the quark moves along the $||$ 
direction, perpendicular to the beam axis which coincides with the 
anisotropy direction.}
\label{fig:sketch}
\end{figure}

Due to collisions, the quark will loose momentum and
the variance of its associated momentum distribution will broaden 
according to \cite{Moore:2004tg}
\bqa
\frac{d}{d\, t} \langle p\rangle &=& - p\, \eta_D(p)\nonumber\\
\frac{d}{d\, t} \langle(\Delta p_{||})^2\rangle &=& \kappa_{||}(p)\nonumber\\
\frac{d}{d\, t} \langle(\Delta p_\perp)^2\rangle &=& \kappa_\perp(p)\nonumber\\
\frac{d}{d\, t} \langle(\Delta p_z)^2\rangle &=& \kappa_z(p),
\label{mastereq}
\eqa
where I choose my coordinate system such that 
$\langle(\Delta p_{||})^2\rangle=(p_{||}-\langle p_{||}\rangle)^2$
is the variance of the momentum distribution in the direction parallel
to the direction of the quark and $\langle(\Delta p_{\perp})^2\rangle,\langle(\Delta p_{z})^2\rangle$
are the variances transverse to the direction of the quark
(see also Fig.\ref{fig:sketch}).

The functions $\eta_D,\kappa_{||},\kappa_\perp,\kappa_z$ which encode
average momentum loss as well as transverse and longitudinal
fluctuations are calculated using kinetic theory. Schematically,
they are given as \cite{Moore:2004tg}
\bqa
\frac{d\langle p\rangle}{d\,t}&=&\frac{1}{2 v}\int_{k,q} |{\mathcal M}|^2 q^0
\left[f(k) (1\pm f(k-q^0))-f(k-q^0)(1\pm f(k))\right]\nonumber\\
\frac{d}{d\, t} \langle(\Delta p_{||})^2\rangle &=&\int_{k,q} |{\mathcal M}|^2
q_{||}^2 f(k) \left[1\pm f(k-q^0)\right]\nonumber\\
\frac{d}{d\, t} \langle(\Delta p_{\perp})^2\rangle &=&\int_{k,q} |{\mathcal M}|^2
q_{\perp}^2 f(k) \left[1\pm f(k-q^0)\right]\nonumber\\
\frac{d}{d\, t} \langle(\Delta p_z)^2\rangle &=&\int_{k,q} |{\mathcal M}|^2
q_z^2 f(k) \left[1\pm f(k-q^0)\right],
\label{schematic}
\eqa
where $v=p/p^0$, $f(k)$ is the gluon (quark) distribution function,
${\mathcal M}$ is the scattering matrix and 
$\int_{k}=\int \frac{d^3 k}{(2\pi)^3}$ (similarly for $q$).  
If the transferred energy
$q^0\simeq {\bf v}\cdot{\bf q}$ is small, one can approximate
\bqa
f(k) \left[1\pm f(k-q^0)\right]-f(k-q^0)
\left[1\pm f(k)\right]&\simeq&-\frac{q^0}{T} f(k) 
\left[1\pm f(k)\right]\nonumber\\
f(k) \left[1\pm f(k-q^0)\right] &\simeq& f(k) \left[1\pm f(k)\right].
\label{simply}
\eqa

Thus, if the quark is non-relativistic or if $q^0\ll T$, 
the coefficient $\kappa_{||}$ can be related
to the energy loss rate $-\frac{dp^0}{dt}$ 
by $\kappa_{||}=-\frac{2T}{v^2} \frac{dp^0}{dt}$. 
This implies that also the other
coefficient functions can be calculated by the appropriate modifications
in the integrand of the energy loss rate.

\subsection{Collisional Energy Loss}

Studies of collisional energy loss of a heavy quark in isotropic
systems have a long
history
\cite{Bjorken:1982tu,Svetitsky:1987gq,
Thoma:1990fm,Braaten:1991we,Romatschke:2004au,Moore:2004tg,
Mustafa:2004dr,Djordjevic:2006tw,Peigne:2005rk,Peshier:2006hi}. 
Assuming a perturbative expansion in powers (and logarithms) of the
strong coupling $\alpha_s=\frac{g^2}{4 \pi}$, 
the leading-order logarithmic contribution to 
the collisional energy loss can be obtained by calculating the matrix
elements in Eqns.~(\ref{schematic}) with HTL propagators and restricting
to soft momentum transfer $q^0<T$ where Eqns.~(\ref{simply}) can be used
\cite{Moore:2004tg}.


There is, however, an alternative way by Thoma and Gyulassy
\cite{Thoma:1990fm} to calculate the collisional
energy loss which is equivalent to calculating the HTL matrix elements
and doing the integrals in Eqns.~(\ref{mastereq}). It is based on the
expression for the energy loss of a quark by interaction with its induced
chromoelectric field,
\beq
\frac{dp^0}{d\, t}={\rm Re} \int d^3x\, {\bf J}^a_{\rm ext}(t,{\bf x})
\cdot {\bf E}^a_{\rm ind}(t,{\bf x}).
\eeq
Here ${\bf J}^a_{\rm ext}=q^a {\bf v}\delta^3({\bf x}-{\bf v}t)$ 
is the current of the quark with color charge $q^a$ and 
velocity $v$ and $E_{\rm ind}$ is the induced electric field. 
In Fourier space $Q=({\bf q},\omega)$, $E_{\rm ind}$ can be written as
\beq
E_{\rm ind}^i(\omega, {\bf q})=i \omega \left(G^{ij}-G^{ij}_0\right)
J_{\rm ext}^j(Q),
\eeq
where $G^{ij}$ and $G^{ij}_0$ are the full and free retarded gluon propagator,
respectively. 

The collisional energy loss for a theory with $N_c$ colors thus becomes
\beq
\frac{dp^0}{d\, t}=-\frac{g^2 (N_c^2-1)}{2 N_c}{\rm Im}
\int \frac{d^4 Q}{(2 \pi)^4} \omega\, v^i
\left(G^{ij}-G^{ij}_0\right) v^j (2\pi) \delta(\omega-{\bf v}\cdot
{\bf q})
\eeq
where the momentum exchange is restricted to $|{\bf q}|<T$ and only
the leading logarithm should be kept.
Because this restriction corresponds to Eqns.~(\ref{simply}), the
fluctuation coefficients may be glossed directly from
Eqns.~(\ref{mastereq}), finding
\bqa
\kappa_{||}&=&\frac{g^2 (N_c^2-1)}{2 N_c}{\rm Im}
\int \frac{d^4 Q}{(2 \pi)^4} \frac{2 T q_{||}^2}{\omega} v^i
\left(G^{ij}-G^{ij}_0\right) v^j (2\pi) \delta(\omega-{\bf v}\cdot
{\bf q})\nonumber\\
\kappa_{\perp}&=&\frac{g^2 (N_c^2-1)}{2 N_c}{\rm Im}
\int \frac{d^4 Q}{(2 \pi)^4} \frac{2 T q_{\perp}^2}{\omega} v^i
\left(G^{ij}-G^{ij}_0\right) v^j (2\pi) \delta(\omega-{\bf v}\cdot
{\bf q})\nonumber\\
\kappa_{z}&=&\frac{g^2 (N_c^2-1)}{2 N_c}{\rm Im}
\int \frac{d^4 Q}{(2 \pi)^4} \frac{2 T q_{z}^2}{\omega} v^i
\left(G^{ij}-G^{ij}_0\right) v^j (2\pi) \delta(\omega-{\bf v}\cdot
{\bf q}).
\label{mydefs}
\eqa

A convenient feature of these Equations is that
all the information about the medium resides in the gluon propagator
$G$. Thus, in this form 
Eqns.~(\ref{mydefs}) are valid both for isotropic as well as
for anisotropic systems.

\subsection{Anisotropic Plasmas}

In the rest frame of a thermal system, particles move in all directions with equal probability. 
However, in the rest frame of a non-equilibrium system, particles might e.g. move predominantly 
in a plane, having an anisotropic probability distribution.
Thus, dropping the assumption of isotropy, there will be at least one
preferred direction in the system, which in the following I will take
to be the $z$-direction. In the context of heavy-ion collisions, one
can identify this direction with the beam-axis along which the system
expands initially. As a model for an anisotropic system, I will assume
that 
the anisotropic quark and gluon distribution functions $f({\bf k})$
are related to the usual (isotropic) Fermi and Bose distributions 
$f_{iso}(|{\bf k}|)$ as \cite{Romatschke:2003ms}
\beq
f({\bf k})=N(\xi)f_{iso}\left(\sqrt{{\bf k}^2+\xi ({\bf k\cdot \bf
e_z})^2}
\right),
\label{aniso}
\eeq
where ${\bf e_z}$ denotes the unit-vector in the z-direction, $\xi$ is
a parameter controlling the strength of the anisotropy and $N(\xi)$ is
a normalization factor which in the following I will set to one.
Note that $\xi=0$ corresponds to the case of an isotropic system.

An anisotropic system by definition does not have one
temperature. However, it may have a dimensionful scale that is related
to the mean particle momentum and which takes over the role of a
``hard'' scale in much the same way the temperature does in a thermal
system. For a quark gluon plasma (even out of equilibrium), the
natural choice seems to be the saturation scale $Q_s$. Since kinetic
theory calculations remain valid even out of equilibrium as long as
there is a separation of scales between hard and soft momentum modes,
one could replace $T$ by $Q_s$ times some coefficient when
translating formulae from an isotropic to an anisotropic system.
For simplicity, however, I will refrain from doing that and keep $T$
as a placeholder of the correct hard scale even for anisotropic systems.

As outlined before, the main difference between calculating the
fluctuation coefficients Eqns.~(\ref{mydefs}) in isotropic and anisotropic
systems, respectively, is the form of the gluon propagator $G$.
While for isotropic systems the propagator in the HTL approximation is
given by\footnote{Here and in the following (unless stated otherwise) 
I use the specific choice
of gauge $A^0=0$; physical observables are not affected by this choice
as they are manifestly gauge-invariant.}
\bqa
G^{ij}_{iso}(Q)&=&\frac{\delta^{ij}-q^i q^j/{\bf q}^2}{{\bf
q}^2-\omega^2+\Pi_T(\omega,{\bf q})}+\frac{q^i
q^j}{\omega^2(\Pi_L(\omega,{\bf q})-{\bf q}^2)}\nonumber\\
\Pi_T(\omega,{\bf q})&=&\frac{m_D^2}{2}
\frac{\omega^2}{{\bf q}^2}\left[1-
\frac{\omega^2-{\bf q}^2}{2 \omega |{\bf q}|} \log{\frac{\omega+|{\bf
q}|}{\omega-|{\bf q}|}}\right]\nonumber\\
\Pi_L(\omega,{\bf q})&=&m_D^2\left[\frac{\omega}{2 |{\bf q}|}
\log{\frac{\omega+|{\bf q}|}{\omega-|{\bf q}|}}-1\right],
\label{isoprop}
\eqa
one finds for systems with an anisotropy of the form Eq.~(\ref{aniso})
within the same approximation \cite{Romatschke:2003ms}
\bqa
G^{ij}(Q)&=&\frac{\delta^{ij}-q^i q^j/{\bf q}^2 
-\tilde{n}^i\tilde{n}^j/{\bf \tilde{n}}^2}{{\bf
q}^2-\omega^2+\alpha(\omega,{\bf q})}+\frac{
\left({\bf q}^2-\omega^2+\alpha(\omega,{\bf q})+\gamma(\omega,{\bf
q})\right) q^i q^j/{\bf q}^2}{
\left({\bf q}^2-\omega^2+\alpha(\omega,{\bf q})+\gamma(\omega,{\bf
q})\right)\left(\beta(\omega,{\bf q})-\omega^2\right)-{\bf q}^2 {\bf
\tilde{n}}^2 \delta^2(\omega,{\bf q})}\nonumber\\
&&+\frac{\left(\beta(\omega,{\bf
q})-\omega^2\right) \tilde{n}^i\tilde{n}^j/{\bf \tilde{n}}^2
-\delta(\omega,{\bf q})\left(q^i \tilde{n}^j+q^j \tilde{n}^i\right)
}{\left({\bf q}^2-\omega^2+\alpha(\omega,{\bf q})+\gamma(\omega,{\bf
q})\right)\left(\beta(\omega,{\bf q})-\omega^2\right)-{\bf q}^2 {\bf
\tilde{n}}^2 \delta^2(\omega,{\bf q})},
\label{genprop}
\eqa
where here ${\bf \tilde n}={\bf e_z}-{\bf q}\, q_z/{\bf q}^2$ and 
the structure functions 
$\alpha,\beta,\gamma,\delta$ are generalizations
of $\Pi_T$ and $\Pi_L$ to anisotropic systems.
In the case of small anisotropy $\xi$, they are given as 
\cite{Romatschke:2003ms}
\bqa
\alpha &=& \Pi_T(\hat{\omega}) + \xi\Bigg[
          {\hat{\omega}^2\over6}(5\hat{q}_z^2 -1)m_D^2 
          -{1\over3}\hat{q}_z^2m_D^2 \nonumber \\
	  && \hspace{3cm} 
	+ {1\over2}\Pi_T(\hat{\omega})\left( (3\hat{q}_z^2 -1) - \hat{\omega}^2(5\hat{q}_z^2 -1)) \right)\Bigg] \; , \nonumber \\
\hat{\omega}^{-2} \beta &=& \Pi_L(\hat{\omega}) + \xi\left[{1\over3}(3\hat{q}_z^2 -1)) m_D^2\right.\nonumber\\
&&\hspace{3cm}\left.
	+ \Pi_L(\hat{\omega}) \left((2\hat{q}_z^2 -1)-{\hat{\omega}^2}(3\hat{q}_z^2 -1)\right)\right] \; , \nonumber \\
\gamma &=& {\xi\over3}(3 \Pi_T(\hat{\omega}) - m_D^2)(\hat{\omega}^2-1)(1-\hat{q}_z^2) \; , \nonumber \\
\delta &=& {\xi\over3k}(4 \hat{\omega}^2 m_D^2+3 \Pi_T(\hat{\omega})(1-4\hat{\omega}^2)) \hat{q}_z\; ,
\label{smallxi}
\eqa
where $\hat{\omega}=\omega/|{\bf q}|$ and $\hat{q}_z=q_z/|{\bf q}|$.
For larger values of $\xi$ one has to resort to numerical evaluations
of $\alpha,\beta,\gamma,\delta$ \cite{Romatschke:2003ms}.

In all cases $m_D$ is the isotropic Debye mass,
\beq
m_D^2=\frac{g^2}{\pi^2}\int_0^\infty dp\ p f_{\rm iso}(|{\bf p}|),
\eeq
which for a thermal system with $N_c$ colors and $N_f$ light quark flavors 
becomes \hbox{$m_D^2=\frac{g^2 T^2 N_c}{3}\left(1+\frac{N_f}{6}\right)$}.

\section{Momentum Broadening}
\label{sec:mombroad}

As can be quickly verified, the general form of the propagator
Eq.~(\ref{genprop}) reduces to the simple form Eq.~(\ref{isoprop}) in
the limit of vanishing anisotropy $\xi\rightarrow 0$. Thus, regardless
whether the system is isotropic or not, the fluctuation coefficients
are found by inserting Eq.~(\ref{genprop}) into Eqns.~(\ref{mydefs}).

The frequency integration is trivial, whereas the integration over
total momentum exchange $|{\bf q}|$ contains a subtlety for
non-vanishing anisotropy $\xi$: since the static limit ($\omega\rightarrow 0$)
of e.g. $\alpha$ is real and negative, the propagator develops a
singularity for space-like momenta $|{\bf q}|$.
Indeed, this singularity signals the presence of instabilities in the
system \cite{Arnold:2002zm,Romatschke:2003ms}.
Since these instabilities correspond to soft gauge modes that
grow nearly exponentially until they reach non-perturbative occupation numbers,
one might dismiss any perturbative approach such as mine as futile.
However, my calculation should be applicable up to the time where the soft
mode occupation number has not grown non-perturbatively large yet. 
Even with fast growing instabilities, this can be a long physical timescale
if either the initial fluctuations are tiny or there is a counter-acting
effect such as the expansion of the system. In the latter case, 
one can even argue that unstable modes do not grow non-perturbatively 
large during plasma lifetimes at RHIC \cite{Romatschke:2006wg}.
Therefore, even though for quantitative results one probably has to resort to
numerical simulations, ignoring the effects of the non-perturbative
soft gauge modes may be not such a bad approximation at least for some period
of time.

As a consequence, here the propagator singularity
is mainly a practical obstacle to calculating physical
observables since this singularity is in general non-integrable, 
as was first recognized by Arnold, Moore
and Yaffe \cite{Arnold:2002zm}. 

However, the fact that $\alpha$ has an imaginary part that vanishes
only linearly in the static limit cures this problem for certain
observables (namely those with numerators that also turn out to vanish
in the static limit), which has been dubbed ``Dynamical Shielding'' 
\cite{Romatschke:2003vc}.

In the formulation I have chosen, Eq.~(\ref{mydefs}), 
there is only one power of the
propagator and consequently the singularity is always integrable.
However, the problem resurfaces in the next integration step, where
for some of the fluctuation coefficients there is an uncanceled
$\omega$ in the denominator of the integral. Unlike the collisional
energy loss, which is finite to full leading-order (LO), dynamical
shielding breaks for at least some of the fluctuation coefficients.
I will show in the following how to calculate these to LL accuracy.

The calculation is divided up into two parts: first I calculate
the ``regular'' contribution within the LL approximation,
which involves only quantities for which dynamical shielding works.
Then I will deal with the contribution which would give the
LO contribution (the constant under the log) in an
isotropic system and show that in anisotropic systems, it gives a
contribution to the LL.

\subsection{Regular contributions}

Limiting oneself to LL approximation, the $|{\bf q}|$
integration in Eq.~(\ref{mydefs}) 
can be done as in Ref.\cite{Romatschke:2003vc},
and one finds e.g.
\bqa
\kappa^{\rm reg}_{\perp}&=&-\frac{g^2 (N_c^2-1)}{2 N_c}
\int \frac{d \Omega_q}{(2 \pi)^3} \frac{4 T
\hat{q}_{\perp}^2}{\hat{\omega}} \frac{1}{1-\hat{\omega}^2}
\log\frac{T}{m_D}{\rm Im}\left[\frac{v^2-\hat{\omega}^2-({\bf
\tilde{n}\cdot v})/\tilde{n}^2}{2(1-\hat{\omega}^2)}\ \alpha 
\right.\nonumber\\
&&\left.
+\frac{(1-\hat{\omega}^2)^2 \beta+\hat{\omega}^2
({\bf \tilde{n}\cdot v})^2/\tilde{n}^2 (\alpha+\gamma)-2 \hat{\omega}
(1-\hat{\omega}^2)({\bf \tilde{n}\cdot v}) \hat{\delta}}
{2 \hat{\omega}^2(1-\hat{\omega}^2)}
\right],
\label{example}
\eqa
where $\hat{q}_{\perp}=q_{\perp}/|{\bf q}|$, 
$\hat{\delta}=\delta/|{\bf q}|$, $\hat{\omega}={\bf v}\cdot{\bf q}/|{\bf q}|$.

I will consider the situation where the quark velocity is perpendicular to
the beam axis, ${\bf v}\cdot{e_z}=0$. Thus I have 
$\hat{\omega}=v \hat{q}_{||}$, ${\bf \tilde{n}\cdot v}=\hat{\omega} 
\hat{q}_z$
and $\tilde{n}^2=1-\hat{q}_z^2$. From the explicit form of the small $\xi$ 
structure functions $\alpha,\beta,\gamma,\delta$ in Eq.~(\ref{smallxi})
it can be seen that their imaginary part always involves at least one
power of $\hat{\omega}$. All the $\hat{\omega}$'s in the denominator of
the integrand in Eq.~(\ref{example}) are thus canceled, showing indeed
that there is no singularity to LO accuracy.

For small anisotropies, the remaining integrations can be done
analytically using Eqns.~(\ref{isoprop},\ref{smallxi}). I find
\bqa
\frac{dp}{d\, t}&=&-\frac{g^2 (N_c^2-1)}{2 N_c} \frac{m_D^2}{8 \pi}
\log\frac{T}{m_D} \left[\frac{1}{v}-\frac{1-v^2}{v^2} {\rm Arctanh}(v)
\right.\nonumber\\
&&\left.\hspace{2cm}+\frac{\xi}{6 v^4}\left(3 v-5 v^3-3 (1-v^2)^2 {\rm Arctanh}(v)\right)\right]
\nonumber\\
\kappa^{\rm reg}_{\perp}&=&\frac{g^2 (N_c^2-1)}{2 N_c} \frac{T m_D^2}{4 \pi}
\log\frac{T}{m_D} \left[\frac{3}{2}-\frac{1}{2v^2}+\frac{(1-v^2)^2}{2
v^3} {\rm Arctanh}(v)
\right.\nonumber\\
&&\left.\hspace{2cm}+\frac{\xi}{24 v^5}\left(-3 v+8 v^3-13v^5+3 (1-v^2)^3 {\rm Arctanh}(v)\right)\right]
\nonumber\\
\kappa^{\rm reg}_{z}&=&\frac{g^2 (N_c^2-1)}{2 N_c} \frac{T m_D^2}{4 \pi}
\log\frac{T}{m_D} \left[\frac{3}{2}-\frac{1}{2v^2}+\frac{(1-v^2)^2}{2
v^3} {\rm Arctanh}(v)
\right.\nonumber\\
&&\left.\hspace{2cm}+\frac{\xi}{24 v^5}\left(-9 v+24 v^3-7v^5+9 (1-v^2)^3 {\rm Arctanh}(v)\right)\right].
\label{regres}
\eqa
As expected, these expression reduce to the known results
\cite{Moore:2004tg} in the isotropic limit ($\xi\rightarrow 0$).

\subsection{Anomalous Contribution}

Let us now concentrate on the naive ``constant under the log'' 
contribution
of Eqns.~(\ref{regres}). To this end, it is illustrative pick out the
first term of the propagator Eq.~(\ref{genprop}), and pluck it into 
Eqns.~(\ref{mydefs}), finding
\beq
\kappa_i\sim
 \int \frac{d^3 q}{(2 \pi)^3} \frac{\hat{q}_i^2}{\hat{\omega}}
\left(v^2-\hat{\omega}^2-\hat{q}^2_z \hat{\omega}^2\right)
{\rm Im}
\frac{q^3}
{{\bf q}^2-\omega^2 + \alpha(\omega,{\bf q})},
\label{starteq}
\eeq
where the index $i$ denotes $i=\perp,||,z$ and I have used the same
notations as in the previous subsection. Denoting this part of the
propagator as $\Delta_A^{-1}={\bf q}^2-\omega^2 + \alpha(\omega,{\bf
q})$ allows me to rewrite
\beq
\kappa_i\sim \int \frac{d^3 q}{(2 \pi)^3} \hat{q}_i^2
m_D^2 f(\hat{q}_z)
\left(v^2-\hat{\omega}^2-\hat{q}^2_z \hat{\omega}^2\right)
q\, \Delta_A\, \Delta_A^* ,
\label{mytestcase}
\eeq
where I introduced $m_D^2 \hat{\omega} f(\hat{q}_z)={\rm Im}\, \alpha$ and 
the $*$ means complex conjugation.
This form now directly involves the squared matrix from
Eqns.~(\ref{schematic}). As discussed before, the static limit of
$\alpha$ is real and negative to leading order in $g$, which produces 
a non-integrable singularity in Eq.~(\ref{mytestcase}) (unless $i=||$
for which $\hat{q}_{||} v=\hat{\omega}$ cuts off the singularity).

However, from the structure of Eq.~(\ref{mytestcase}), this singularity
is reminiscent of so-called pinching singularities, which are usually
due to incomplete resummations of the propagator. Indeed, I find it is plausible
that $\alpha$ has a non-vanishing imaginary part in the static
limit at order\footnote{See the appendix for details.} 
$O(g^3)$, which allows one to integrate 
\bqa
\int dq\, q^3 |\Delta_A|^2&=&
\frac{1}{2(1-\hat{\omega}^2)^2} \left[\frac{{\rm Re}\,\alpha}{{\rm
Im}\,\alpha} {\rm Arctan}\left({\frac{{\rm
Re}\,\alpha+q^2(1-\hat{\omega}^2)}{{\rm
Im}\,\alpha}}\right)\right.\nonumber\\
&&\left.
\hspace{2cm}+\frac{1}{2}\log\left(|\alpha|^2+2 {\rm Re}\, \alpha\, q^2
(1-\hat{\omega}^2) +q^4 (1-\hat{\omega}^2)^2
\right)
\right].
\label{con15}
\eqa
The second term in this equation gives a contribution to the 
LL for $q=T\gg m_D$ which has been accounted for already in the
previous subsection. The other term, however, now involves ${\rm Im}\,
\alpha$ rather than the $\hat{\omega}$ of Eq.~(\ref{starteq}) 
in the denominator, which because of the $O(g^3)$ contribution behaves
as 
\beq
\lim_{\hat{\omega}\rightarrow 0+} {\rm Im}\, \alpha\sim m_D^2
\left(\hat{\omega}+c g\right),
\eeq
where $c$ is just a number. Thus, the singularity at
$\hat{\omega}\rightarrow 0+$ is cut-off by this higher-order
resummation in the self-energy. However, this also entails that the
naive LO correction gives a contribution to the
LL instead, as I will show in the following.

First note that for any nonzero $\hat{\omega}$, the first term on the
r.h.s. of Eq.~(\ref{con15}) is finite also for $c=0$ and thus only
contributes to LO. In order to extract a
potential LL contribution, one may thus take
$\hat{\omega}\rightarrow 0^+$ everywhere {\em except} for the $\hat{\omega}$ in the
denominator, which one has to replace by 
$\hat{\omega}\rightarrow \hat{\omega}+ c\, g$.
For simplicity, I do this already in Eq.~(\ref{starteq}), using
\footnote{
This $\delta$-function is not to be confounded with the structure function $\delta$!}
\beq
\lim_{\hat{\omega}\rightarrow 0+} {\rm Im} \Delta_A\rightarrow + \pi
\delta\left({\bf q}^2 + {\rm Re}\, \alpha\right).
\label{improp}
\eeq

As can be verified, for small anisotropies $\xi$ the leading 
contribution to $\kappa_i$ is entirely given by this part of the
propagator\footnote{This is because the structure function
$\hat{\delta}^2$ in the denominator may be dropped
to leading order in $O(\xi)$. However,
this result is specific to choosing the quark
velocity perpendicular to ${\bf e_z}$, so that 
${\bf \tilde{n}\cdot v}\rightarrow 0$ in the static limit.}.
The integral over $\hat{\omega}$ then gives $\log{(g)}$ together with
some finite LO contribution that I ignore.

Thus, upon identifying $\log{(g)}=-\log{\frac{T}{m_D}}$, 
I find for the anomalous contributions to $\kappa_i$ 
\bqa
\kappa_\perp^{\rm anom}&=&-\frac{g^2 (N_c^2-1)}{2 N_c} \frac{T m_D^2}{4
\pi} \log\frac{T}{m_D} \frac{\xi\, v}{12}\nonumber\\
\kappa_z^{\rm anom}&=&-\frac{g^2 (N_c^2-1)}{2 N_c} \frac{T m_D^2}{4
\pi} \log\frac{T}{m_D} \frac{\xi\, v}{4},
\eqa
such that together with the regular contributions Eq.~(\ref{regres}), the
full LL fluctuation coefficients in the small
anisotropy limit are given by
\bqa
\kappa_{\perp}&=&\frac{g^2 (N_c^2-1)}{2 N_c} \frac{T m_D^2}{4 \pi}
\log\frac{T}{m_D} \left[\frac{3}{2}-\frac{1}{2v^2}+\frac{(1-v^2)^2}{2
v^3} {\rm Arctanh}(v)
\right.\nonumber\\
&&\left.\hspace{2cm}+\frac{\xi}{24 v^5}\left(-3 v+8 v^3-13v^5-2 v^6+3 (1-v^2)^3 {\rm Arctanh}(v)\right)\right]
\nonumber\\
\kappa_{z}&=&\frac{g^2 (N_c^2-1)}{2 N_c} \frac{T m_D^2}{4 \pi}
\log\frac{T}{m_D} \left[\frac{3}{2}-\frac{1}{2v^2}+\frac{(1-v^2)^2}{2
v^3} {\rm Arctanh}(v)
\right.\nonumber\\
&&\left.\hspace{2cm}+\frac{\xi}{24 v^5}\left(-9 v+24 v^3-7v^5-6 v^6+9
(1-v^2)^3 {\rm Arctanh}(v)\right)\right].
\label{fullres}
\eqa

\subsection{Larger Anisotropies}

The relevant integrals of the previous subsections may also be
evaluated numerically for arbitrary $\xi$ 
using techniques from \cite{Romatschke:2003ms,Romatschke:2003vc}.
The results are shown in Table \ref{tab:one}.
For sufficiently small $\xi$, the results turn out to coincide with
the analytic results Eqns.~(\ref{fullres}).

From Table \ref{tab:one}, it can be seen that 
the anomalous contribution makes up only a few percent at small $\xi$,
while it becomes more important for larger anisotropies.
Indeed, at larger $\xi$, besides Eq.~(\ref{improp}) then also the other parts of the propagator 
contribute,
further enhancing the anomalous contributions. For the total $\kappa_z$, where the anomalous
contribution is stronger, this manifests itself by a decrease as a function of $v$
for larger $\xi$, whereas the regular contribution would have had the inverse trend.

Another trend that is apparent from Table \ref{tab:one} is that
the ratio $\kappa_z/\kappa_\perp$ for a given anisotropy $\xi$
decreases for increasing velocity $v$. Put differently, the higher
the quark's momentum, the more circular its associated jet shape.

\begin{table}
\begin{tabular}{|c|cccc|cccc|cccc|}
\hline
&&\quad $\xi=$&$\!\!\!\!\!\!\!\!\!1$&&&\quad $\xi=$
&$\!\!\!\!\!\!\!\!\!10$&&& 
$\xi=$& $\!\!\!\!\!\!\!\!\!100$&\\
v & $\kappa_{\perp}$& ($\kappa_{\perp}^{\rm reg}$) & $\kappa_z$ & 
($\kappa_z^{\rm reg}$) & 
$\kappa_{\perp}$ & ($\kappa_{\perp}^{\rm reg}$) & $\kappa_z$ & 
($\kappa_z^{\rm reg}$) &  
$\kappa_{\perp}$ & ($\kappa_{\perp}^{\rm reg}$) & $\kappa_z$ & ($k_z^{\rm reg}$)\\
\hline
\hline
\ 0.05\  & \ 	0.50 &  (0.50) & 0.98 & (1.00) 	&
		0.23 &	(0.23) & 1.45 & (1.54)  &
		0.08 &	(0.08) & 1.72 & (1.84)	\\
\ 0.15\  & \ 	0.50 &  (0.51) & 0.97 & (1.00)	&
 		0.22 &	(0.23) & 1.44 & (1.54)	&
		0.07 &	(0.08) & 1.69 & (1.84)	\\
\ 0.25\  & \ 	0.51 &  (0.51) & 0.97 & (1.01) 	&
		0.22 &	(0.24) & 1.40 & (1.54)	&
		0.07 &	(0.08) & 1.63 & (1.84)	\\
\ 0.35\  & \ 	0.52 &  (0.53) & 0.96 & (1.02) 	&
		0.22 &	(0.25) & 1.35 & (1.55)	&
		0.07 &	(0.08) & 1.55 & (1.84)	\\
\ 0.45\  & \ 	0.53 &  (0.55) & 0.96 & (1.03)	&
		0.22 &	(0.26) & 1.31 & (1.55)  &
		0.07 &	(0.09) & 1.47 & (1.85)	\\
\ 0.55\  & \ 	0.54 &  (0.57) & 0.96 & (1.05)	&
		0.23 &	(0.27) & 1.26 & (1.55)	&
		0.07 &	(0.09) & 1.40 & (1.85)	\\
\ 0.65\  & \ 	0.56 &  (0.60) & 0.97 & (1.07)	&
		0.24 &	(0.29) & 1.22 & (1.56)	&
		0.07 &	(0.10) & 1.32 &	(1.85)	\\
\ 0.75\  & \ 	0.58 &  (0.64) & 0.98 & (1.10)	&
		0.25 &	(0.31) & 1.17 &	(1.57)	&
		0.07 &	(0.11) & 1.24 &	(1.85)	\\
\ 0.85\  & \ 	0.60 &  (0.69) & 1.00 & (1.14)	&
		0.27 &	(0.34) & 1.13 & (1.58)	&
		0.08 &	(1.12) & 1.16 &	(1.85)	\\
\ 0.95\  & \ 	0.64 &  (0.75) & 1.02 & (1.19)	&
		0.30 &	(0.38) & 1.09 & (1.59)	&
		0.09 &	(0.14) & 1.08 &	(1.85)	\\
\hline
\end{tabular}
\caption{The coefficients multiplying 
$\frac{g^2 (N_c^2-1)}{2 N_c} \frac{T m_D^2}{4\pi} \log\frac{T}{m_D}$
in $\kappa_\perp,\kappa_z$ for larger values of the system anisotropy
$\xi$, evaluated numerically. For convenience, also the regular
contributions are given explicitly.}
\label{tab:one}
\end{table}

\section{Discussion}
\label{sec:discussion}

Both from the analytic results Eqns.~(\ref{fullres}) as well from the
numeric evaluation given in Table \ref{tab:one}, one finds that for 
anisotropic systems $\kappa_z/\kappa_\perp$ is always larger than one.
In other words, a charm quark jet in an anisotropic quark-gluon plasma
will generically experience more broadening along the longitudinal
direction than in the reaction plane.
Assuming an initial jet profile with 
$\langle(\Delta p_\perp)_0^2\rangle=\langle(\Delta p_z)_0^2\rangle$, 
the ratio $\kappa_{z}/\kappa_\perp$ 
can roughly be associated with the ratio
of jet correlation widths in azimuth $\langle\Delta \phi\rangle$ and rapidity 
$\langle\Delta\eta\rangle$ as 
\beq
\kappa_{z}/\kappa_\perp\simeq \frac{\langle\Delta \eta\rangle}{\langle\Delta \phi\rangle}.
\label{correq}
\eeq
Therefore, the calculation in the previous sections generically
predicts this ratio to be larger than one in anisotropic
systems. 

\begin{figure}
\begin{center}
\includegraphics[width=\linewidth,viewport=0 450 575 700]{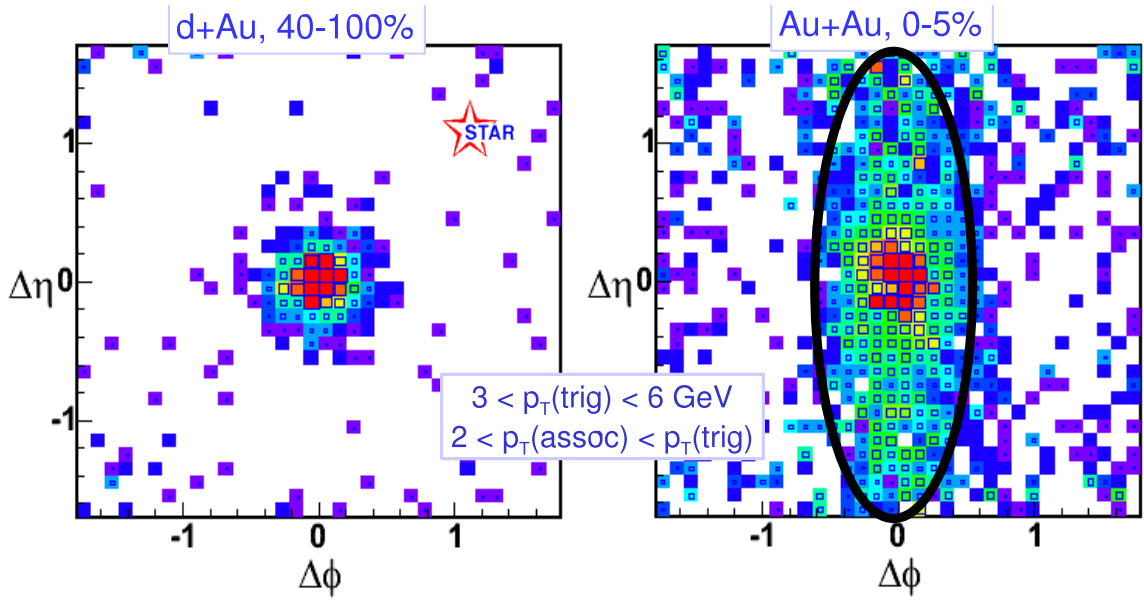}
\end{center}
\caption{Dihadron correlation functions in azimuth $\Delta \phi$ and 
space-time rapidity $\Delta \eta$, for d+Au and central Au+Au at 
$\sqrt{s}=200$ GeV (figure courtesy J. Putschke (STAR), Proceedings
of Hard Probes 2006).
For comparison, on the right plot I show an ellipse with eccentricity 
$e\simeq \frac{\sqrt{8}}{3}$ which corresponds to a 
ratio $\kappa_z/\kappa_\perp\sim 3$.}
\label{fig:ptcorr}
\end{figure}

Due to the approximations in my calculation (heavy quark, 
leading-log, only collisional broadening), 
many effects that will change both $\kappa_z$ 
and $\kappa_\perp$ for real jets have been ignored. However, 
inclusion of other effects should not change my result on the \emph{ratio} of
$\kappa_z$ and $\kappa_\perp$ qualitatively, since $\kappa_z=\kappa_\perp$ 
in isotropic systems. Thus, $\kappa_z/\kappa_\perp>1$ 
in anisotropic systems should hold true also for real jets.
Interestingly, recent measurements of the dihadron
correlation functions by STAR \cite{Jacobs:2005pk} 
for Au+Au collisions 
at $\sqrt{s}=200$ GeV indeed seem to imply much broader correlations 
in $\Delta \eta$ than in $\Delta\phi$.
In Fig.\ref{fig:ptcorr}, STAR data for these correlation is compared
to a general ellipsoidal shape with eccentricity 
$e\simeq \frac{\sqrt{8}}{3}$, which according to Eq.~(\ref{correq})
I would associate with a ratio $\kappa_z/\kappa_\perp\sim 3$.

In other words, a charm quark jet with a ``mean'' 
momentum $p\simeq4.6$GeV (or $v\simeq 0.95$) and a ``mean'' 
system anisotropy $\xi\simeq 10$ should experience broadening 
roughly consistent with the ellipsoidal shape in
Fig.\ref{fig:ptcorr}, where ``mean'' here is
referring to a mean over the whole system evolution.
To get more quantitative, one would have to calculate the momentum
broadening dynamically according to Eqns.~(\ref{mastereq}), which 
can e.g. be done within a viscous hydrodynamics simulation
\cite{Muronga:2001zk,Heinz:2005bw,Baier:2006um}.
However, note that as a general feature of
anisotropic plasmas, one seemingly obtains jet
shapes that can be much broader in rapidity than in azimuth. 
Other effects distorting the shape of jets in an isotropic plasma,
namely flow effects \cite{Armesto:2004pt}, 
probably cannot account for such dramatic asymmetries
unless invoking extreme flows.

Interestingly, momentum-binned data of the dihadron
correlations from STAR seem to show a trend towards more 
circular jet shapes 
for higher trigger momentum, 
which I also found in section \ref{sec:mombroad}.C.
Precise data could thus provide further tests for calculations, maybe helping
to constrain the system anisotropy.

In view of this, a finite system anisotropy could be a natural
explanation for these broad rapidity correlations 
in central Au+Au collisions at $\sqrt{s}=200$ GeV.
Thus, turning the argument around,
it may be that the sizable ratio $\langle\Delta
\eta\rangle/\langle\Delta \phi\rangle$  is
an indication that the plasma created at RHIC is after all
\emph{not} in equilibrium during a sizable fraction of its lifetime,
calling into question the validity of the 
strongly advocated ``perfect fluid'' picture.

Clearly, in order to shed more light onto 
this issue, many caveats on the application of 
my result to experimental data have to be addressed.
Among other things,
a calculation of the full LO correction
as well as results for light quarks, gluons and the inclusion of
radiation effects in anisotropic plasmas would be on the wish-list.
Nevertheless, I hope that I was able to highlight the feasibility and 
potential value of re-calculating experimentally interesting observables 
for anisotropic plasmas.

To summarize, I have calculated the rates governing momentum
broadening for a heavy quark to leading-log accuracy in an anisotropic
quark-gluon plasma. The potential singularity, naively plaguing the
constant under the log of these rates is shown to be cured 
after resummations of the NLO gluon self-energy, eventually 
contributing to the leading log. I expect
this procedure to render finite also other observables that are 
affected by these sort of singularities, making their calculation
feasible also in anisotropic plasmas. 
The rates found in this calculation suggest jet shapes that are
elongated along the rapidity direction, which could explain
recent STAR data and thus point towards the possibility of
non-equilibrium phenomena at RHIC.

\acknowledgments

I would like to thank Rob Pisarski for suggesting to look 
into this problem and Adrian Dumitru, Peter Jacobs, Anton Rebhan and 
Larry Yaffe for fruitful discussions.
Moreover, I am indebted to Rudolf Baier for his clever 
suggestions that turned out to work so well.
Finally, I want to thank BMBF06BI102 for financial support.

\begin{appendix}
\section{NLO Gluon Self-Energy}

\begin{figure}
\begin{center}
\includegraphics[width=0.8\linewidth]{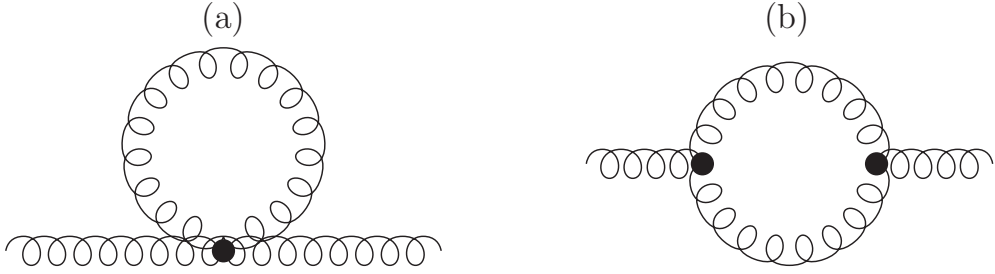}
\end{center}
\caption{Diagrams contributing to the gluon self-energy}
\label{fig:gluon}
\end{figure}

I will be interested in the static limit of the imaginary part of the
gluon self-energy to $O(g^3)$. Following Braaten and Pisarski
\cite{Braaten:1990it}, this can be calculated by dressing the
propagators and vertices of the contributing diagrams.

A further simplification occurs due to the static limit: following
Rebhan \cite{Rebhan:1993az}, only static loop momenta have to be
resummed and no fermionic loops contribute. 
Since the ghost self-energy vanishes at leading order, 
the only contribution to the gluon self-energy at $O(g^3)$ 
in the static limit are thus the two diagrams shown in 
Fig. \ref{fig:gluon}.

Of these, the tadpole diagram (shown in Fig.\ref{fig:gluon}(a)) 
with bare vertices contributes \cite{Rebhan:1993az}
\beq
\delta \Pi^{\mu \nu}=-g^2 N_c \int \frac{d^3 k}{(2\pi)^3}
\lim_{\omega\rightarrow 0}\left[g^{\mu \nu}
(G^{\alpha}_{\alpha}(\omega,{\bf k}) -G^{\alpha}_{(0)\, \alpha}(\omega,{\bf k}))-(G^{\mu\nu}(\omega,{\bf k})-G_{(0)}^{\mu \nu}(\omega,{\bf k}))
\right],
\eeq
where $G$ and $G_0$ denote the dressed and free propagator, respectively.
The dressed propagator in covariant gauges 
can be found by inverting the relation
\beq
g_{\mu \nu} K^2 +(\frac{1}{\rho}-1) K_\mu K_\nu + \Pi_{\mu \nu} =
G^{-1}_{\mu \nu},
\label{geninvpropdef}
\eeq
where $\rho$ is the gauge parameter and $\Pi^{\mu \nu}$ is the gluon
self-energy to $O(g^2)$ which for isotropic as well as anisotropic
systems can be expressed \cite{Romatschke:2003ms} 
by the functions $\alpha,\beta,\gamma,\delta$ introduced in the main text.

For isotropic systems, inversion of Eq.~(\ref{geninvpropdef}) is
trivial and one indeed recovers the known result 
\cite{Blaizot:2000fc} for the tadpole contribution to $\delta \Pi$.

For anisotropic systems, the inversion of Eq.~(\ref{geninvpropdef}) is a
little more involved but it is readily performed by e.g. a symbolic
manipulation package on a computer. 
Taking the imaginary part of $\delta \Pi^{\mu \nu}$, the only
non-vanishing contribution after performing the static limit 
comes from the singularities of the propagator.
These, however, give just the delta-functions considered in the main
text, c.f. Eq.~(\ref{improp}).

In the small-$\xi$ limit, the integrals may then be done analytically,
and one finds
\bqa
\lim_{\omega\rightarrow 0+}{\rm Im}\, \delta \alpha(\omega,{\bf k}) &=& \frac{-g^2 N_c m_D T}{16 \pi} 
\left(2-\hat{k}_z^2\right)
\sqrt{\frac{\xi}{3}}
\left(2-\frac{1}{4} \log{\left(3+2 \sqrt{2}\right)}\right)
\nonumber\\
\lim_{\omega\rightarrow 0+}{\rm Im}\, \delta \gamma(\omega,{\bf k}) &=& \frac{g^2 N_c m_D T}{8 \pi} 
\hat{k}_z^2
\sqrt{\frac{\xi}{3}}
\left(2-\frac{1}{4} \log{\left(3+2 \sqrt{2}\right)}\right).
\label{NLOres}
\eqa

Therefore, there is a gauge-invariant, non-vanishing $O(g^3)$
contribution to the imaginary part of the structure functions 
in the static limit. Since I find it highly improbable that the
corresponding contribution from the second diagram in Fig. 
\ref{fig:gluon}(b) or anisotropic vertex corrections
cancel this contribution \emph{exactly}, I will
refrain from evaluating also these contributions, and instead 
take Eqns.~(\ref{NLOres}) as indication that the \emph{whole} $O(g^3)$
contribution does not vanish. 

However, both diagrams will have
to be recalculated at non-vanishing frequency when aiming for the
full LO contribution
to the fluctuation coefficients $\kappa_i$.

\textbf{Note added:} In thermal equilibrium systems (which are necessarily
isotropic) the KMS condition requires the imaginary part of 
the self-energy to vanish in the static limit \cite{LeBellac}.
Out of equilibrium, the imaginary part of the self-energy 
has to be an odd function of frequency \cite{Chou:1984es}, 
but this does not seem to preclude
a discontinuity at vanishing frequency. To show unambiguously
whether there is a non-vanishing imaginary part of the self-energy
in the static limit for non-equilibrium system, one has to 
calculate the above diagrams with both resummed propagators and vertices, 
which is interesting and doable, but beyond the scope of this work.

\end{appendix}

\end{document}